\documentclass[12pt,letterpaper,copyedit]{article} % single column, double spaced
\usepackage[tablesfirst,notablist,nomarkers]{endfloat} %% float figs. to back

\usepackage{ol2}
\usepackage[draft]{hyperref}
\usepackage{amsmath}

\begin{document}

%\twocolumn[ %% activate for two-column option

\title{Sub-hertz frequency stabilization of a commercial diode laser}

%% For REVTeX it is possible to automate superscript and e-mail callouts with the superscriptaddress option; see REVTeX4 documentation.

\author{Y. N. Zhao,$^1$ J. Zhang,$^1$ J. Stuhler,$^2$ G. Schuricht,$^2$ F. Lison,$^2$ Z. H. Lu,$^1$ and L. J. Wang$^{1,*}$}

\address{$^1$Max Planck Institute for the Science of Light\\
and Institute of Optics, Information and Photonics, University Erlangen\\
G\"unther-Scharowsky-Str. 1, Building 24, 91058 Erlangen, Germany\\}
\address{$^2$TOPTICA Photonics AG\\
Lochhamer Schlag 19, 82166 Graefelfing, Germany}
\address{$^*$Corresponding author: Lijun.Wang@mpl.mpg.de}

\begin{abstract}We report ultra-stable locking of a commercially available extended cavity diode laser to a vibration-insensitive high finesse Fabry-Perot cavity. A servo bandwidth of 2 MHz is demonstrated. The absolute stability of the diode laser after locking is measured with a three-cornered-hat method. The resulting Allan deviation reaches a level of $2.95\times10^{-15}$ at 1 s, corresponding to only 0.93 Hz linewidth, even without vibration isolation of the reference cavity.
\end{abstract}


% ] %% activate for two-column option

\noindent Ultranarrow linewidth lasers are essential for optical clocks and high precision laser spectroscopy. Due to their broad tuning ranges, ease of use, and lower cost, diode lasers are particularly suitable for these tasks. To reach an ultranarrow linewidth, the laser is generally servo locked to a high finesse Fabry-Perot cavity through Pound-Drever-Hall (PDH) technique \cite{Telle,Ohtsu,Hilico,Notcutt,Ludlow,Alnis}. Unlike monolithic Nd:YAG lasers \cite{Webster1}, diode lasers can still have significant frequency noise at Fourier frequencies above 1 MHz. As a consequence, it is much more difficult to design a servo loop for an extended cavity diode laser than for an intrinsically quiet Nd:YAG laser. Specially designed servo electronics is generally required to suppress the diode laser linewidth down to Hz level, limiting this technology to very few advanced labs.

In this Letter, we report locking of a commercial extended cavity diode laser (ECDL) to a vibration-insensitive high finesse Fabry-Perot cavity with a commercial wide-band electronic servo amplifier. A locking bandwidth of 2 MHz is demonstrated. Using the three-cornered-hat method, the absolute stability of the diode laser is measured to be $2.95\times10^{-15}$ at 1 s, corresponding to only 0.93 Hz linewidth. This performance level is achieved even without vibration isolation of the reference cavity.

\begin{figure}[htb]
\centering
\includegraphics[width=14cm]{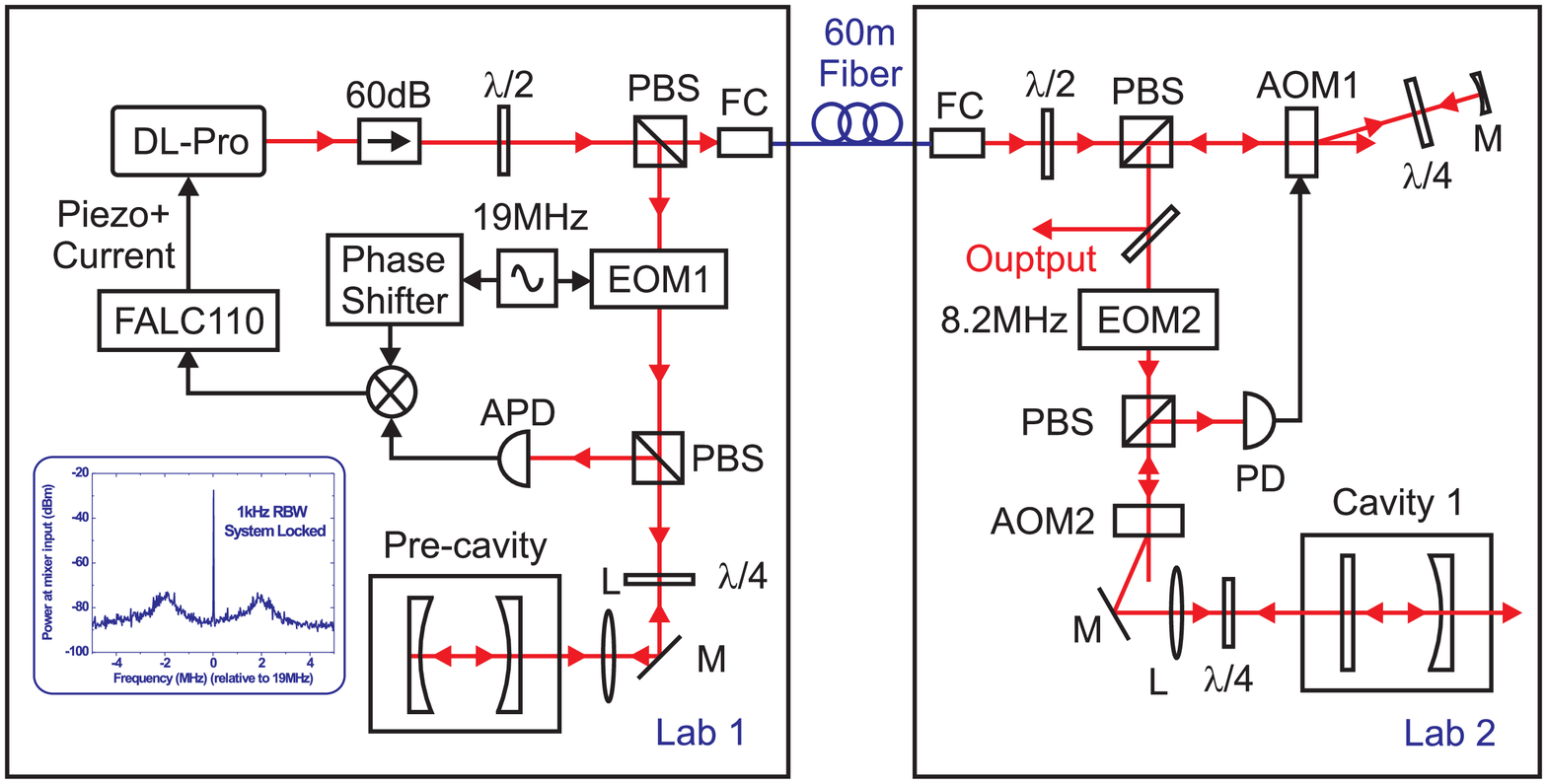}
\caption{(Color online) Experimental setup for diode laser locking. EOM, electro-optic modulator; AOM, acousto-optic modulator; $\lambda/2$, half wave plate; $\lambda/4$, quarter wave plate/ PBS, polarization beam splitter; APD, avalanche photo diode; PD, photodiode; FC, fiber coupler; M, mirror; L, lens. Inset, The reflection signal from the pre-cavity recorded by a spectrum analyzer upon laser locking.}
\label{fig1}
\end{figure}

The experimental setup is shown in Fig.~\ref{fig1}. A commercial ECDL (TOPTICA DL pro 940) is placed in one lab (lab 1), while a high finesse reference cavity (cavity 1) is placed in another acoustically isolated lab (lab 2). The diode laser is tuned to 946 nm (the 4th harmonic at 236.5 nm is used to probe the 5s$^2$ $^1$S$_0$ -- 5s5p $^3$P$_0$ clock transition of the $^{115}$In$^+$ optical frequency standard) and emits 30 mW of power. A 5 $\mu$s delayed self-heterodyne measurement results in a short term linewidth of the DL-pro laser of $\sim100$ kHz. The laser frequency can be modulated by applying a voltage to a piezoelectric transducer (PZT) which moves the ECDL's optical grating. This is possible at frequencies from DC to $3.5$ kHz before a mechanical resonance of the grating holder is hit. High frequency laser modulation can be performed via an active electric circuit in the laser head which allows to alter the laser diode current from DC to 60 MHz (3 dB bandwidth). The frequency response of the diode laser with respect to current modulation depends on the individual laser diode type and typically falls of at a few MHz. The reference cavity is designed to be vibration-insensitive \cite{Nazarova,Webster3,Zhao}. This is done by making square ``cutouts'' on the bottom of the cylindrical spacer and supporting the cavity at four Airy points. For such a cavity, the important parameter is the ratio of the cavity length change d$L$ over seismic acceleration d$g$. Here, the cavity acceleration sensitivity d$L/$d$g$ is found to be less than $10^{-14}$ s$^2$ using a finite element analysis method. The cavity spacer and the optically bonded mirror substrates are made of ultra-low-expansion glass (ULE). The cavity has a finesse of $2.4\times10^6$ measured by cavity ring-down time. It is placed on a home-made stainless steel base with four supporting points. The whole assembly is placed inside a temperature-stabilized ultrahigh vacuum chamber. The temperature of the vacuum chamber is stabilized at $27\ ^{\circ}$C with a root-mean-square (RMS) fluctuation of less than 1 mK. To reduce the influence of acoustic noise, the vacuum chamber is surrounded by plastic foams containing a layer of lead septum. The cavity and other optical components sit on top of a 10 cm thick breadboard without any vibration isolation.

The diode laser light is sent through a 60 m long single mode fiber from lab 1 to lab 2. This will introduce a one-way loop delay time of 300 ns. Since maximally achievable control bandwidth is limited by the overall loop delay time \cite{Mor}, for a loop bandwidth of a few MHz, the loop delay has to be under 50 ns. To solve this problem, we pre-stabilize the diode laser to another simple pre-cavity. The pre-cavity is also made of ULE, and has a finesse of only $73\ 000$. The pre-cavity is placed in a vacuum chamber for isolation from air temperature and pressure fluctuations. 

The pre-locking is performed through PDH technique. A small amount of the diode laser output is sent through a 19 MHz electro-optical modulator (EOM1), and is coupled into the pre-cavity. The cavity reflection signal is used to generate the error signal, which is sent to a commercial fast analog linewidth controller (FALC) servo amplifier (TOPTICA FALC 110). Within the FALC, the error signal is split into two regulator branches. A slow integrator branch acts on the laser grating PZT to cancel out long term drifts and the fast branch (signal delay times $\sim15$ ns) is used to control the laser diode current. For fastest settings, the FALC transfer function shows a phase delay of $<45^{\circ}$ at 35 MHz and $\sim90^{\circ}$ at 50 MHz with the 3 dB bandwidth of 100 MHz. 
%Its circuit is split into two amplifier channels. One fast circuit branch is used to drive high-bandwidth laser current modulator circuits. Another low-speed integrating amplifier with high DC gain is used to drive PZT. The overall transfer function of the fast circuit branch is set by configuring three different lag-lead filter stages and a lead-lag filter. The lag-lead filters act as the integrators, and perform the main functions of the servo loop since the frequency noise of diode lasers commonly decreases with frequency. The lead-lag filter acts as the differentiator, and allows for phase advance in the control loop at high frequencies. All filter stages can be disabled individually depending on circumstances. This allows for advanced control of the phase. The gain slopes of the individual filter stages are changed by setting the DIP switches, while the gain levels of the filter stages are usually kept fixed.

By optimizing FALC settings, we achieve a pre-locking servo bandwidth of $\sim 2$ MHz, as shown in the inset of Fig.~\ref{fig1}. The reflected signal from the pre-cavity, after pre-locking, is recorded by a spectrum analyzer with a 1 kHz resolution bandwidth. The center peak corresponds to the modulation frequency of EOM1. Shoulders appear around this peak, indicating a locking bandwidth of 2 MHz. The linewidth of the diode laser is thus reduced from $\sim 100$ kHz level to a few hundreds Hz, limited by the mechanical stability of the pre-cavity. About 5 mW of the remaining diode laser light is subsequently sent through the 60 m fiber to lab 2. The light from the fiber output double-passes through an acouto-optic modulator (AOM1), which is used to bridge the frequency difference between the two closest TEM$_{00}$ modes of pre-cavity and cavity 1, and is also used for PDH locking to cavity 1. The light after double-passing AOM1 is separated into two parts. One part is used for the three-cornered-hat measurement to determine the absolute stability of the diode laser. The other part of the light is mode-matched into cavity 1 for PDH locking, with the reflected signal as the error signal. The error signal is used to control AOM1 to complete the feedback loop. Since frequency noise at Fourier frequencies above kHz has been suppressed by the pre-locking, the requirement on the second stage servo electronics bandwidth is greatly reduced. We just need a simple, low bandwidth ($< 15$ kHz) servo amplifier for the locking. AOM2 is used as an optical isolator to prevent standing waves on the photo detector, PD.

\begin{figure}[htb]
\centerline{\includegraphics[width=10cm]{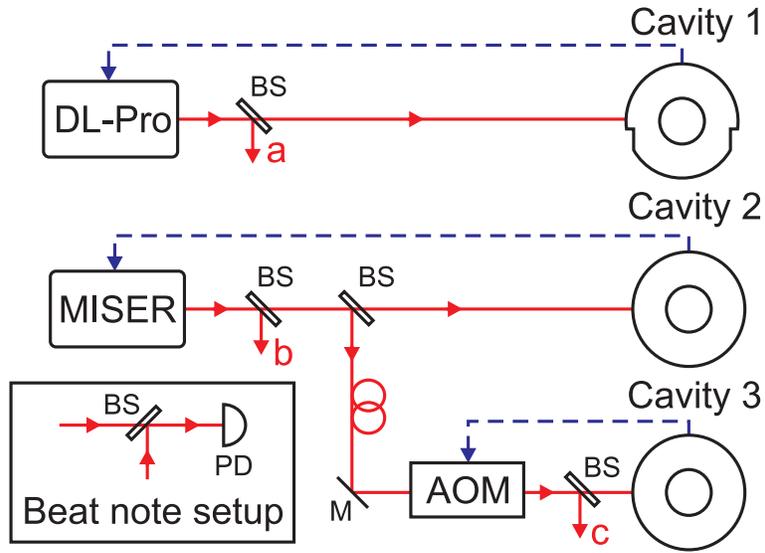}}
\caption{(Color online) Schematic setup for the three-cornered-hat measurement. BS, beam splitter; AOM, acousto-optic modulator. The light beams at a, b, and c points are sent to the respective beat note setups as shown in the inset at the left side.}
\label{fig2}
\end{figure}

To measure the absolute stability of the diode laser, we perform a three-cornered-hat measurement \cite{Zhao,Liu1}. As shown in Fig.~\ref{fig2}, The output from a monolithic isolated end-pumped ring Nd:YAG laser (MISER) is split into two parts, and independently locked to two other reference cavities, cavity 2 and cavity 3 \cite{Liu2}. Both cavity 2 and cavity 3 are put on top of active vibration isolation stages. Together with the diode laser output, three independent beams are formed, and three cross-beat frequencies are obtained at approximately 304 MHz, 392 MHz, and 88 MHz for cavity 1-2, 1-3, and 2-3, respectively. We then down-convert the beat frequencies to the kHz level and use low-pass filters to clean up the signals. The three down-converted signals are recorded by three frequency counters (Agilent 53132A) for gate time longer than 1 s, and by three fast analog-digital converters (Gage CompuScope 12400) for gate time shorter than 1 s. 

\begin{figure}[htb]
\centerline{\includegraphics[width=14cm]{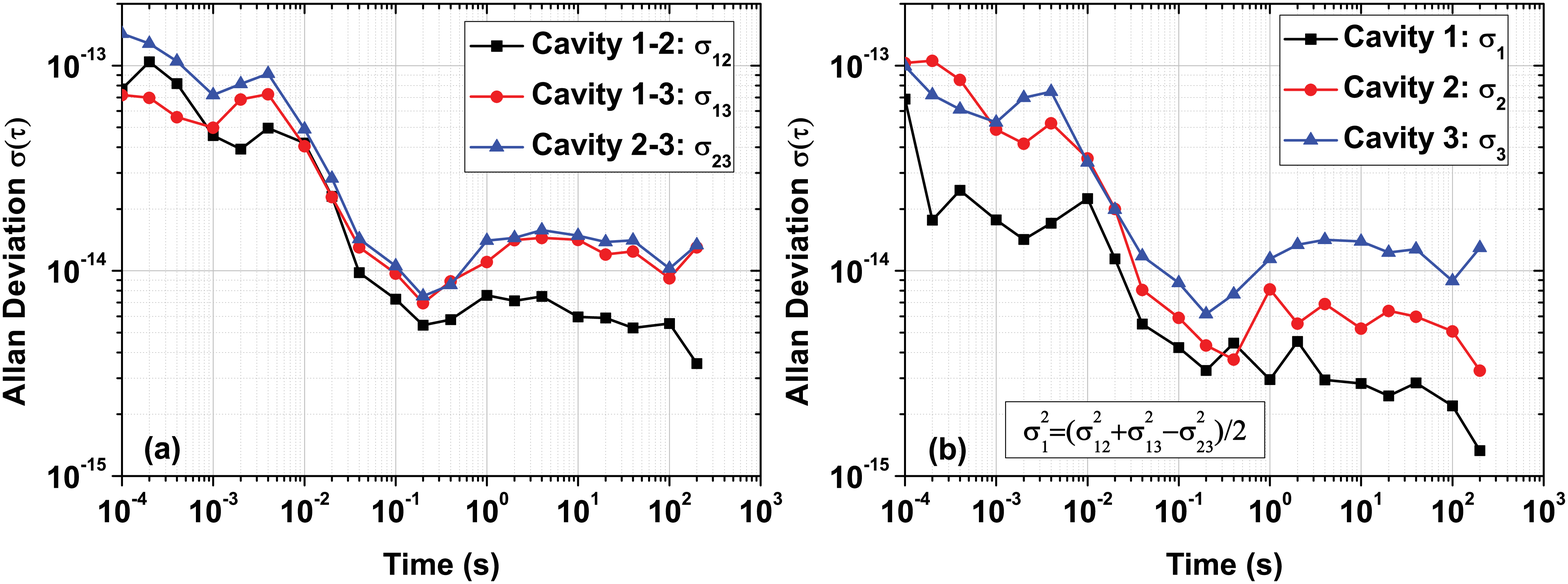}}
\caption{(Color online) (a) The Allan deviations of the beat frequencies between three reference cavities. The linear frequency drifts of approximately 1 Hz/s are removed during data processing. (b) The calculated Allan deviations of the individual reference cavities.}
\label{fig3}
\end{figure}

The calculated Allan deviations of the three data sets are shown in Fig.~\ref{fig3}(a). Linear frequency drifts of approximately 1 Hz/s are removed during data processing. The frequency stabilities of individual cavities can then be calculated with the three-cornered-hat method. The final results are shown in Fig.~\ref{fig3}(b). It shows that cavity 1 has the best frequency stability over the entire measurement time from 100 $\mu$s to 200 s, even without active vibration isolation. An absolute stability of $2.95\times10^{-15}$ of cavity 1 is measured at 1 s, corresponding to only 0.93 Hz in linewidth.

In conclusion, we demonstrate locking of a commercially available ECDL to a vibration-insensitive high finesse Fabry-Perot cavity with a commercially available wide-band electronic servo loop. A locking bandwidth of 2 MHz has been achieved. Even higher bandwidth can be obtained by rearranging the placements of the diode laser and the pre-cavity to further reduce the loop delay time. The absolute frequency stability of the diode laser is measured with a modified three-cornered-hat method to remove correlation effects. We also design and implement a vibration-insensitive reference cavity. As a consequence, even without active vibration isolation, a high stability of $2.95\times10^{-15}$ at 1 s is achieved. The result is better than those of two normal cavities with active vibration isolation.

Due to the complex servo electronics design, the sub-hertz frequency stabilizations of diode lasers demonstrated previously \cite{Notcutt,Ludlow,Alnis} are limited to only a few advanced labs. With this work, we demonstrate that sub-hertz locking of diode lasers based entirely on commercially available components is feasible. We believe that a modular design, ultranarrow linewidth diode laser covering a broad wavelength range can greatly aid further development of precision laser spectroscopy.

%\begin{figure}[htb]
%\centerline{\includegraphics[width=8cm]{fig2.eps}}
%\caption{The reflection signal from the pre-cavity recorded by a spectrum analyzer upon laser locking. The center peak corresponds to the modulation frequency of EOM1. Shoulders appear around this peak, indicating a locking bandwidth of 2 MHz.}
%\label{fig2}
%\end{figure}

\pagebreak %To balance final column

%\bibliographystyle{ol} %Remove and replace with bbl-file contents for submission
%\bibliography{Refs} %Remove and replace with bbl-file contents for submission

\begin{thebibliography}{99}
\newcommand{\enquote}[1]{``#1''}

\bibitem{Telle} H. R. Telle, ``Narrow Linewidth Laser Diodes with Broad Continuous Tuning Range,'' Appl. Phys. B {\bf 49,} 217 (1989).
\bibitem{Ohtsu} M. Ohtsu, K. Nakagawa, M. Kourogi, and W. Wang, ``Frequency control of semiconductor lasers,'' J. Appl. Phys. {\bf 73,} R1 (1993).
\bibitem{Hilico} L. Hilico, D. Touahri, F. Nez, and A. Clairon, ``Narrow-line, low-amplitude noise semiconductor laser oscillator in the 780 nm range,'' Rev. Sci. Instrum. {\bf 65,} 3628 (1994).
\bibitem{Notcutt} M. Notcutt, L. -S. Ma, J. Ye, and J. L. Hall, ``Simple and compact 1-Hz laser system via an improved mounting configuration of a reference cavity,'' Opt. Lett. {\bf 30,} 1815 (2005).
\bibitem{Ludlow} A. D. Ludlow, X. Huang, M. Notcutt, T. Zanon-Willette, S. M. Foreman, M. M. Boyd, S. Blatt, and J. Ye, ``Compact, thermal-noise-limited optical cavity for diode laser stabilization at $1\times10^{-15}$,'' Opt. Lett. {\bf 32,} 641 (2007).
\bibitem{Alnis} J. Alnis, A. Matveev, N. Kolachevsky, Th. Udem, and T. W. H\"{a}nsch, ``Subhertz linewidth diode lasers by stabilization to vibrationally and thermally compensated ultralow-expansion glass Fabry-P\'{e}rot cavities,'' Phys. Rev. A {\bf 77,} 053809 (2008).
\bibitem{Webster1} S. A. Webster, M. Oxborrow, and P. Gill, ``Subhertz-linewidth Nd:YAG laser,'' Opt. Lett. {\bf 29,} 1497 (2004).
%\bibitem{Webster2} S. A. Webster, M. Oxborrow, S. Pugla, J. Millo, and P. Gill, ``Thermal-noise-limited optical cavity,'' Phys. Rev. A {\bf 77,} 033847 (2008).
\bibitem{Nazarova} T. Nazarova, F. Riehle, and U. Sterr, ``Vibration-insensitive reference cavity for an ultra-narrow-linewidth laser,'' Appl. Phys. B {\bf 83,} 531 (2006).
\bibitem{Webster3} S. A. Webster, M. Oxborrow, and P. Gill, ``Vibration insensitive optical cavity,'' Phys Rev. A {\bf 75,} 011801(R) (2007).
\bibitem{Zhao} Y. N. Zhao, J. Zhang, A. Stejskal, T. Liu, V. Elman, Z. H. Lu, and L. J. Wang, ``A vibration-insensitive optical cavity and absolute determination of its ultrahigh stability,'' http://arxiv.org/abs/0904.0865.
\bibitem{Mor} O. Mor and A. Arie, ``Performance analysis of Drever-Hall laser frequency stabilization using a proportional + integral servo,'' IEEE J. Quant. Electron. {\bf 33}, 532 (1997).
\bibitem{Liu1} T. Liu, Y. N. Zhao, V. Elman, A. Stejskal, and L. J. Wang, ``Characterization of the absolute frequency stability of an individual reference cavity,'' Opt. Lett. {\bf 34}, 190 (2009).
\bibitem{Liu2} T. Liu, Y. H. Wang, R. Dumke, A. Stejskal, Y. N. Zhao, J. Zhang, Z. H. Lu, L. J. Wang, Th. Becker, and H. Walther, ``Narrow linewidth light source for an ultraviolet optical frequency standard,'' Appl. Phys. B. {\bf 87}, 227 (2007).
%\bibitem{Premoli} A. Premoli and P. Tavella, ``A revisited three-cornered hat method for estimating frequency standard instability,'' IEEE Trans. Instrum. Meas. {\bf 42}, 7 (1993).

\end{thebibliography}

\end{document}